\documentclass[twocolumn,superscriptaddress,aps]{revtex4}

\usepackage{graphicx}
\usepackage{dcolumn}
\usepackage{bm}
\usepackage{multirow}

\begin{document}

\title{A spin- and angle-resolving photoelectron spectrometer}

\author{M.~H.~Berntsen}
\affiliation{Materials Physics, KTH Royal Institute of Technology, S-16440 Kista, Sweden}
\author{P.~Palmgren}
\affiliation{MAX-lab, P.O. Box 118, S-22100 Lund, Sweden}
\author{M.~Leandersson}
\affiliation{MAX-lab, P.O. Box 118, S-22100 Lund, Sweden}
\author{A.~Hahlin}
\author{J.~\AA hlund}
\affiliation{VG Scienta AB, P.O. Box 12120, SE-75015 Uppsala, Sweden}
\author{B.~Wannberg}
\affiliation{VG Scienta AB, P.O. Box 12120, SE-75015 Uppsala, Sweden}
\affiliation{BW Particle Optics AB, P.O. Box 55, SE-82222 Alfta, Sweden}
\author{M.~M\aa nsson}
\affiliation{Laboratory for Neutron Scattering, ETH Zurich \& Paul Scherrer Institut, CH-5232 Villigen PSI, Switzerland}
\affiliation{Laboratory for Synchrotron and Neutron Spectroscopy, EPFL, CH-1015 Lausanne, Switzerland}
\author{O.~Tjernberg}
\email{oscar@kth.se}
\affiliation{Materials Physics, KTH Royal Institute of Technology, S-16440 Kista, Sweden}

\date{\today}

\begin{abstract}
A new type of hemispherical electron energy analyzer that permits angle and spin resolved photoelectron spectroscopy has been developed. The analyzer permits standard angle resolved spectra to be recorded with a two-dimensional detector in parallel with spin detection using a mini-Mott polarimeter. General design considerations as well as technical solutions are discussed and test results from the Au(111) surface state are presented.
\end{abstract}

\pacs{}
                             
\maketitle
\section{Introduction}
Angle-resolved photoelectron spectroscopy (ARPES) is a very powerful technique for the study of electronic structures in solids. The technique has improved dramatically during the past two decades with the rapid progress in photon sources as well as electron energy analyzers. The introduction of two-dimensional detectors was a very important step in the development of the analyzers, which permitted parallel acquisition in energy as well as angle. As a result, it has become a routine measurement to determine the energy dispersion along a reciprocal space line. In a standard ARPES experiment, the energy and emission angle of the photoemitted electron is determined. Depending on sample, geometry and light polarization, the photoemitted electron may also carry spin information. Determining the electron spin is, however, a complicated task and there are at present no available techniques that permit parallel acquisition in angle and energy together with spin analysis. The common implementation of spin- and angle-resolved photoelectron spectroscopy (SARPES) therefore relies on selecting a particular energy and angle for spin analysis. Since the spin analysis is also a very inefficient process in itself, it becomes a very time-consuming process to map out the spin-resolved energy dispersion. In practice one would therefore first like to perform normal ARPES in order to rapidly determine orientation and energy dispersion, and then perform spin resolved measurements at selected points in reciprocal space. In this article, we present the design and implementation of a spectrometer capable of performing such measurements.

\section{System overview}
\begin{figure*}
\includegraphics[width=\textwidth]{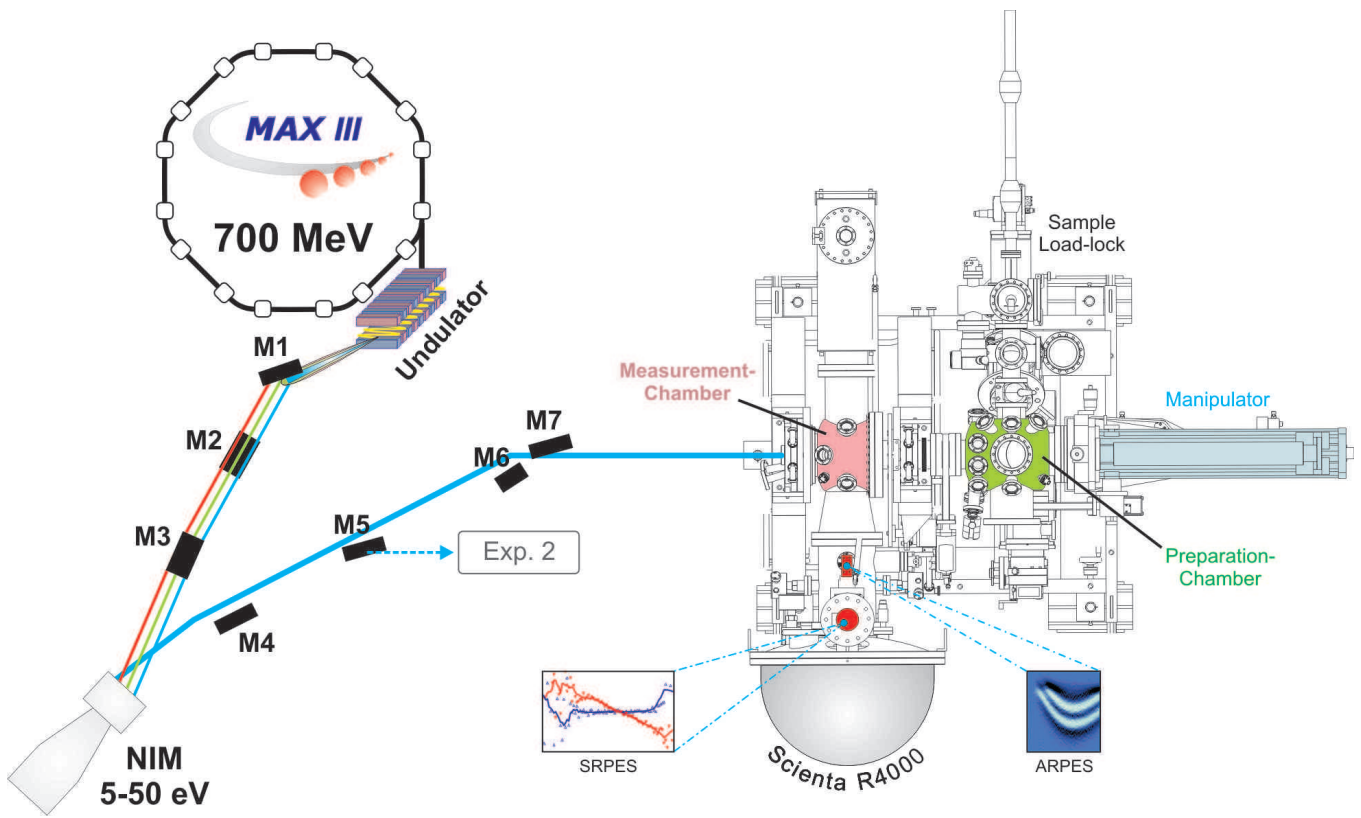}
\caption{Schematic overview of the NIM beamline and the spin-ARPES end station (I3) at MAX III.}
\label{fig:system}
\end{figure*}
The system described in this article is located at the MAX III synchrotron of the MAX-laboratory in Lund, Sweden. MAX III is a third generation 700~MeV synchrotron dedicated to low photon energy experiments. The light source of the I3 beam line is a 2~m eliptically polarizing undulator optimized for photon energies in the 5-50~eV range. A high resolution 6.65~m eagle-type normal incidence monochromator (NIM) with three gratings provides a maximum resolving power of $\approx 10^5$. The schematic layout of the beam line is displayed in the left part of Fig.~\ref{fig:system}. As shown in the figure, the beam line has two branch lines. One branch is intended for user endstations (labeled Exp.~2) while the other branch is permanently equipped with the spin-ARPES endstation. An overview drawing of the endstation is shown in the right part of Fig.~\ref{fig:system}. Referring to this figure, the measurement chamber is seen to the left and the beam enters this chamber from above at angle of 16 degrees from the horizontal plane. The measurement chamber is equipped with a hemispherical electron energy analyzer with a double detector system consisting of one spin- and one two-dimensional photoelectron detector and is rotatable around the horizontal axis. As a result, it is possible to cover all emission angles without changing the incidence angle of the photons. To the right of the measurement chamber, separated by a gate valve, is the preparation chamber which is equipped with tools for standard preparation and characterization techniques such as ion bombardment, Low Energy Electron Diffraction (LEED) and Auger Electron Spectroscopy (AES). Sample positioning is achieved through a horizontally placed sample manipulator that reaches both the preparation and measurement chambers. The manipulator has five degrees of freedom and permits resistive sample heating as well as liquid Helium cooling. In order to permit fast introduction of samples as well as sample storage the endstation is also equipped with a load-lock and a transfer chamber. The transfer chamber is also used for connecting to a nearby Molecular Beam Epitaxy (MBE) system, allowing samples to be transferred in vacuum between the two systems.

\section{Analyzer}
\begin{figure}
\includegraphics[width=8.5cm]{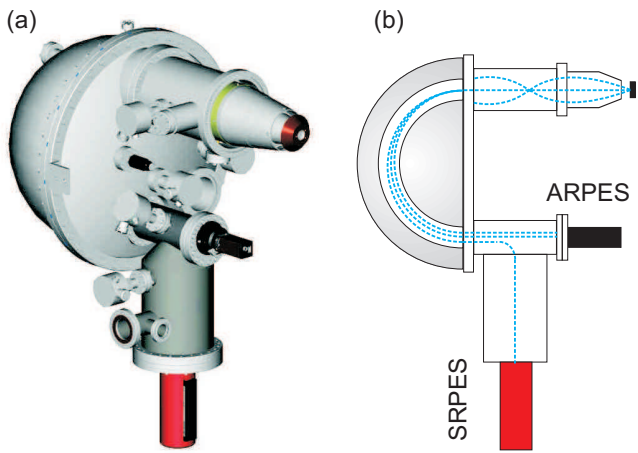}
\caption{(a) A three-dimensional view of the hemispherical analyzer including the CCD camera for angle-resolved photoelectron spectroscopy (ARPES), the spin-transfer lens and the Mott detector used for spin-resolved photoelecton spectroscopy (SRPES). (b) Schematic overview of electron trajectories through the analyzer.}
\label{analfig}
\end{figure}

\begin{figure}
\includegraphics[width=8.5cm]{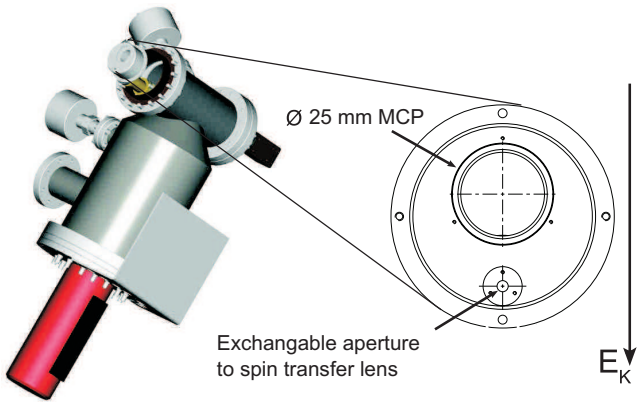}
\caption{The transfer lens and the Mott detector shown together with a detailed view of the placement of the two-dimensional detector and the transfer lens entrance aperture.}
\label{transfig}
\end{figure}

The key component of the system is the analyzer which is based on the VG Scienta R4000 analyzer. The general analyzer layout is shown in
Fig.~\ref{analfig}. It consists of the lens, the hemispheres, a two-dimensional detector setup, a spin-transfer lens and a mini-Mott spin detector. The lens, slits, apertures and hemispheres are identical to those of the standard R4000 analyzer. However, the standard 40 mm diameter two-dimensional detector of the R4000 has been replaced by an off center 25~mm two-dimensional detector and a circular aperture for the spin-transfer lens as demonstrated in Fig.~\ref{transfig}. The two-dimensional detector is a circular Micro Channel Plate (MCP) assembly followed by a fluorescent screen which is monitored by a digital camera system. All in all, the setup is very similar to the normal 40~mm diameter detector setup of the R4000 except for the off center placement and an increased camera distance (400~mm rather than 140~mm) to accommodate the transfer-lens. The reduced detector size is expected to give a corresponding reduction in count rate of about 60\% for a given angular dispersion and pass energy. However, the energy and angular resolution of the two-dimensional detector is unaffected by the reduced size and it is still possible to select different angular dispersion modes with acceptance angles up to $\pm15^{\circ}$.

As displayed in Fig.~\ref{transfig}, the aperture of the spin-transfer lens is located on the center-line in angular direction of the two-dimensional detector but displaced along the energy axis. The nominal aperture of the spin-transfer lens is 4~mm but inserts can be used to reduce the aperture size. Calculated energy and angular resolutions for different apertures, pass energies and angular modes are given in Table~I \& II. Changing the aperture involves venting the system and removing the transfer lens. This is a complicated and time consuming operation but since the energy and angular resolution contributions from the Mott detector can be selected through different pass energies and angular modes it is not expected that there will be any need to change the aperture.

Once the electrons enter the transfer lens they are accelerated from the pass energy up to $\approx$~500~eV and then deflected 90$^{\circ}$. The deflection serves two purposes. First, it rotates the spin quantization axes of the polarimeter so that the spin polarization can be measured both along and perpendicular to the lens axis. Second, it acts as an energy filter and suppresses any unwanted secondary electrons. After the deflection, the electrons are further accelerated and enters the mini-Mott polarimeter where they strike the scattering target at a kinetic energy of 25~keV.
The mini-Mott polarimeter is of the Rice University type \cite{Burnett94} with a 120$^{\circ}$ scattering geometry, four individual 25~mm diameter channeltrons and a thorium target at 25~kV potential. This polarimeter has an expected Sherman function of $S_{eff}=0.17$ \cite{Burnett94}. The choice of this type of polarimeter was based on several factors. One is that the electron optical acceptance of the polarimeter has to be sufficiently large. Even though this is not an requirement for angle resolved modes, large acceptance is needed when running the analyzer in transmission mode. In case of the latter, and for higher pass energies, the electron optical acceptance needs to be in the order of 100~mm$^2\cdot$sr$\cdot$eV. This requirement is not a problem for a mini-Mott or a diffuse scattering polarimeter \cite{Klebanoff93} but is a disqualifying factor for the spin-LEED polarimeter \cite{sawler91,ghir99}. The repeated regeneration of target surfaces required for the diffuse scattering and spin-LEED polarimeters was also considered a serious drawback in a system based at a synchrotron were users have limited time and reliability is important. Therefore, the choice fell on the mini-Mott design, which is also sufficiently compact to fit on the rotatable analyzer where a high voltage Mott would have been too bulky and heavy.

\begin{table}
\begin{tabular*}{\columnwidth}{@{\extracolsep{\fill}}  cccc}
\multicolumn{4}{c}{Energy resolution (meV)} \\ \hline
Pass- & \multicolumn{3}{c}{Aperture} \\
energy (eV) & 2mm & 3mm & 4mm \\ \hline
1 & 5 & 7.5 & 10\\
2&10&15&20\\
5&25&37.5&50\\
10&50&75&100\\
20&100&150&200\\
50&250&375&500\\
100&500&750&1000\\
200&1000&1500&2000 \\ \hline \\
\end{tabular*}
\caption{The different energy resolution contributions from the spin detector as a function of pass energy and aperture. For larger analyzer slits the analyzer energy resolution must also be taken into account.}
\end{table}

\begin{table}
\begin{tabular*}{\columnwidth}{@{\extracolsep{\fill}}  cccc}
\multicolumn{4}{c}{Angular resolution (degrees)} \\ \hline
Angular & \multicolumn{3}{c}{Aperture} \\
mode & 2mm & 3mm & 4mm \\ \hline
7 & 0.75 & 1.12 & 1.5\\
15 & 1.5 & 2.25 & 3\\
30 & 3 & 4.5 & 6\\ \hline \\
\end{tabular*}
\caption{The different angular resolutions obtainable with the spin detector as a function of angular mode and aperture.}
\end{table}

\section{Results}
In order to demonstrate the functionality of the system and verify the assumed Sherman function for the polarimeter, the surface state of Au(111) has been studied. As shown in previous photoemission experiments \cite{LaShell96,Reinert01,Nicolay01} the parabolic surface state of Au(111) is split into two branches. Theoretical studies \cite{LaShell96,Petersen00} indicate that the splitting is an electron spin related phenomena and that the surface state is fully spin polarized. The observed splitting is a consequence of the broken inversion symmetry which occurs at the surface where the bulk crystal is terminated. The broken symmetry allows spin-orbit interaction to lift the degeneracy of the surface state producing a splitting of the free-electron-like parabolic dispersion into two branches with energies \cite{Petersen00}
\begin{eqnarray}
 E(k)=\frac{\hbar^2k^2}{2m}\pm \alpha k,
 \label{eq:FE}
\end{eqnarray}
where $\hbar$ is the Planck constant, $m$ the effective electron mass, $k$ the electron in-plane momentum and $\alpha$ a so called strength parameter. In a simple 2D free-electron model the strength parameter $\alpha$ is proportional to the gradient of the potential which confines the electrons to the surface. However, more realistic models of the atomic fields are needed in order to reproduce the magnitude of the splitting \cite{Petersen00,LaShell96,Osterwalder04}. The fact that theory predicts 100\% polarization makes the surface state a suitable candidate for initial calibration purposes, which explains why this particular test subject has been chosen for the work presented here.
\begin{figure}
	\centering
		\includegraphics[width=8.5cm]{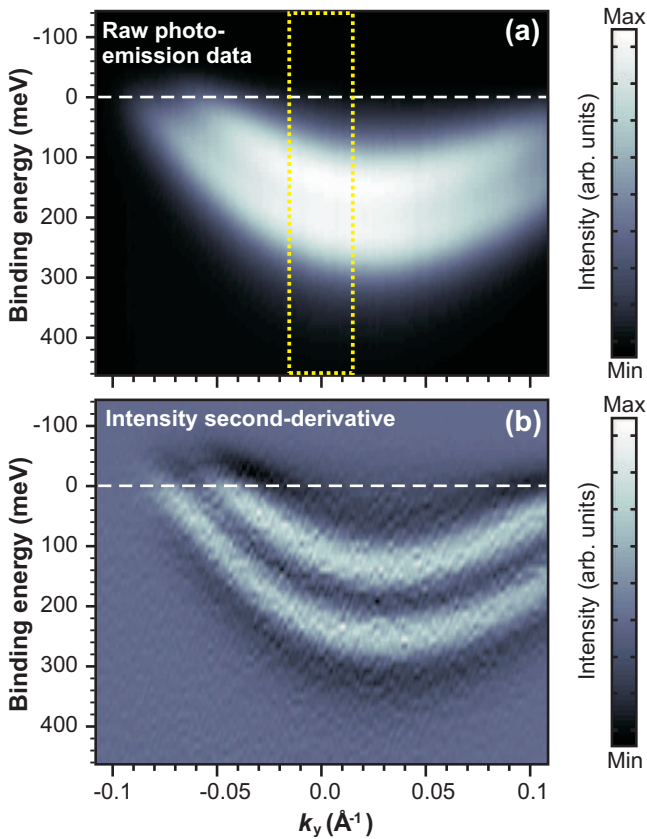}
	\caption{(a) Room-temperature photoemission spectrum (raw data) of Au(111) surface state measured at $\theta=6.8^{\circ}$ from normal emission using linearly polarized light at $h\nu=10$ eV. Dotted rectangle (yellow) indicates $k$-interval corresponding to the $1.5^{\circ}$ window for which the spin-resolved data is acquired. Intensities are given by the colorbar to the right. (b) Second derivative of the intensity with respect to energy for the spectrum showed in (a). Two distinct branches of the free-electron-like parabola are clearly resolved.}
	\label{fig:spectrafig}
\end{figure}

The experiment was conducted on an Au(111) single crystal mounted on a sample holder which allows control over both polar ($\theta$) and azimuthal ($\phi$) angles. The measurement was performed at ultra-high vacuum (UHV) conditions (base pressure $<2\cdot 10^{-10}$~Torr) after preparing the surface by repeated cycles of Ar-ion sputtering and annealing at 1~kV and $550^{\circ}$C respectively. Surface reconstruction and quality was checked with LEED. Room-temperature data were collected using linearly polarized light with photon energy, $h\nu=10$~eV. Fig.~\ref{fig:spectrafig}(a) shows the raw data spectrum acquired with the MCP detector parallel to the $\Gamma-M$ direction ($k_y$-axis) at a polar angle $\theta=6.8^{\circ}$ from normal emission. The rotation in polar angle is necessary in order to move away from $(k_x,k_y)=(0,0)$, since there is no spin split at this point in $k$-space. As seen in the figure, the spectrum is shifted towards positive $k_y$-values and the parabolas have their minima located at $k_y=0.026$ \AA$^{-1}$, which indicates that the sample is slightly tilted, i.e. a slight rotation about the $k_x$-axis. The two branches reach the Fermi level at $k_F=-0.059$ \AA$^{-1}$ (upper) and $k_F=-0.091$ \AA$^{-1}$ (lower) resulting in $\Delta k_{F}=0.032$ \AA$^{-1}$, in agreement with previously obtained values \cite{LaShell96,Osterwalder04}. In Fig.~\ref{fig:spectrafig}(b) we display the second-derivative of the intensity with respect to binding energy of the spectrum in Fig.~\ref{fig:spectrafig}(a). Even though the measurement was performed at room-temperature the two dispersive branches of the free-electron-like surface state are clearly distinguishable. The broadening of the features in the spectrum in Fig.~\ref{fig:spectrafig} is mainly due to thermal broadening. Since the photon source has an energy resolution $<1$~meV a total ultimate energy resolution of $\sim$1~meV for angle-resolved measurements is expected.
\begin{figure}
	\centering
		\includegraphics[width=8.5cm]{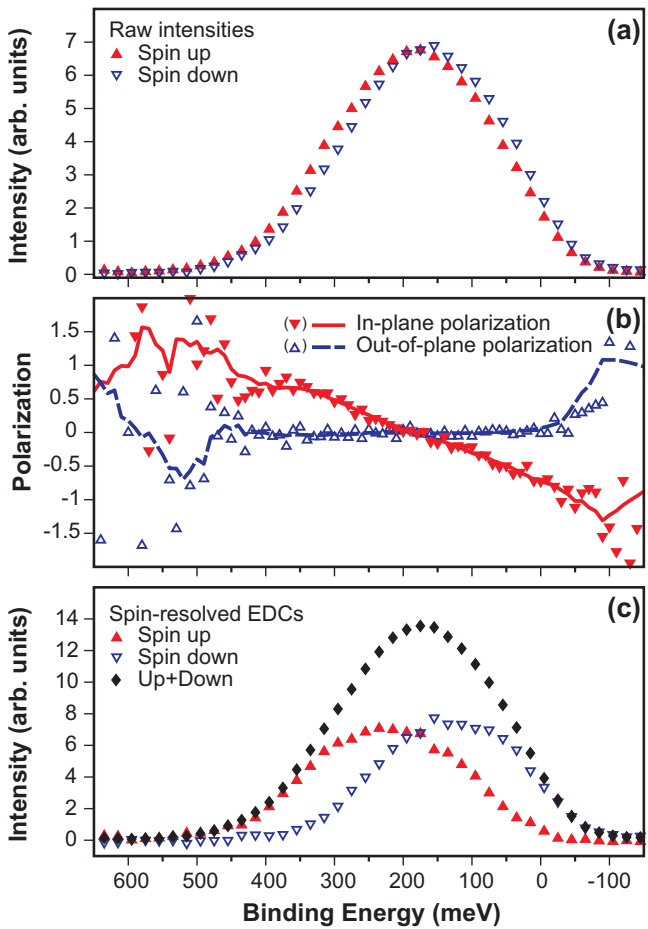}
	\caption{(a) Raw data from spin-up and spin-down channels of the mini-Mott detector after background subtraction and normalization. Filled (red) triangles correspond to spin up, open (blue) triangles to spin down. (b) Measured polarization as a function of binding energy. Filled triangles correspond to the in-plane polarization calculated from the intensities in (a) assuming a Sherman function $S_{eff}=0.17$. Open triangles indicate the out-of-plane polarization using the same $S_{eff}$. Solid and dotted lines are running averages. (c) Spin resolved spectra of Au(111) surface state measured at room-temperature and $6.8^{\circ}$ away from normal emission with energy- and angular resolutions of 100~meV and $1.5^{\circ}$ respectively.}
	\label{fig:spinfig}
\end{figure}

Generally, the spin-polarization of electrons with energy $E$ is defined by the asymmetry
\begin{eqnarray}
 P(E)=\frac{n^{\uparrow}(E)-n^{\downarrow}(E)}{n^{\uparrow}(E)+n^{\downarrow}(E)},
 \label{eq:Asym1}
\end{eqnarray}
where $n^{\uparrow(\downarrow)}(E)$ is the spin up (down) electron energy distribution. In this case, since the spin sensitivity of the polarimeter is $<1$ the polarization of the surface state can be found by dividing the measured asymmetry between the upper and lower channel by the effective Sherman function $S_{eff}$ assuming a negligible instrumental asymmetry, i.e. 
\begin{eqnarray}
 P(E)=\frac{I^{\uparrow}(E)-I^{\downarrow}(E)}{[I^{\uparrow}(E)+I^{\downarrow}(E)]S_{eff}}.
 \label{eq:Asym2}
\end{eqnarray} 
Here, $I^{\uparrow(\downarrow)}$ represent the intensities in the up- and down channels respectively. When analyzing the data, the instrumental asymmetry was set to zero by normalizing the intensities from the up and down channels. This can be justified by the fact that the surface state is fully polarized and hence should result in equal intensities in the upper and lower channel provided that the channels are properly calibrated.

The spin-resolved measurement was performed using the Angular 15 mode with pass energy 20~eV and a spin-transfer lens aperture of 2~mm. The resulting energy- and angle resolution for the spin detector is 100~meV and $1.5^{\circ}$ respectively (see Table~I and II).
Consequently, the spin-data presented in Fig.~\ref{fig:spinfig} were acquired within a 0.032 \AA$^{-1}$ ($1.5^{\circ}$) wide $k$-window centered at $k=0$, indicated by dotted, yellow rectangle in Fig.~\ref{fig:spectrafig}(a).
Fig.~\ref{fig:spinfig}(a) shows the resulting intensities from the spin up and down channels after subtraction of integrated background and normalization. From these intensities, the polarization shown in Fig.~\ref{fig:spinfig}(b) is determined using the expected Sherman function $S_{eff}=0.17$ \cite{Burnett94}. Given that the Au(111) surface state is 100\% polarized we expect the in-plane component of the polarization to approach +(-)1 at lower(higher) binding energies. From Fig.~\ref{fig:spinfig}(b) we see that this is indeed the case, except for some oscillations at the upper and lower limits of the spectrum which results from low statistics. In addition, the out-of-plane component of the polarization, determined using the intensities from the left and right channels of the polarimeter, is zero for all energies, which is consistent with a spin oriented in the plane. Worth noting is that previous spin-resolved measurements \cite{Osterwalder04} on the Au(111) surface state only reached $\approx40\%$ polarization in spite of the theoretical predictions of a fully polarized state. Hence, the results presented here does not only confirm the assumed Sherman function $S_{eff}=0.17$ and the proper function of the instrument but also provides strong support for the predictions regarding the spin structure of the Au(111) surface state. The spin-resolved EDCs presented in Fig.~\ref{fig:spinfig}(c) are obtained using the determined in-plane polarization $P(E)$ and the average spin-integrated intensity $I(E)$ from all four channels of the spin detector through the relation
\begin{eqnarray}
 I_{\pm}(E)=[1\pm P(E)]I(E),
 \label{eq:SpinIntes}
\end{eqnarray} 
where +/- indicates spin up and down respectively. Although partly overlapping, the two spin-peaks are distinguishable and their peak positions are separated by approximately 100~meV, a value in agreement with previous measurements \cite{LaShell96,Osterwalder04}.

\section{Summary}
In conclusion, the instrument presented in this paper offers the possibility of measuring the spin orientation of photoemitted electrons by the use of a mini-Mott polarimeter. The additional ability to perform conventional ARPES measurements is a clear advantage since it allows the energy dispersion and sample orientation in $k$-space to be rapidly determined before proceeding with the spin measurement. Hence, contamination sensitive surfaces, i.e. samples with short life-time, and materials with very complex band-structure can more easily be studied. The initial test performed on the Au(111) surface state verifies that the instrument works properly and confirms the assumed Sherman function of the polarimeter as well as the complete spin polarization of the surface state.

\section{Acknowledgements}
The instrumental development described here was made possible through grants from the K\&A Wallenberg foundation as well as the Swedish Research Council (VR).

\bibliography{references}

\end{document}